\documentclass[doublecol]{epl2}

\title{Making new connections towards cooperation in the prisoner's dilemma game}
\shorttitle{Making new connections towards cooperation in the prisoner's dilemma game}

\author{Attila Szolnoki,\inst{1} Matja{\v z} Perc\inst{2} \and Zsuzsa Danku\inst{3}}
\shortauthor{Attila Szolnoki \textit{et al.}}

\institute{\inst{1}Research Institute for Technical Physics and Materials Science, P.O. Box 49, H-1525 Budapest, Hungary\\
\inst{2}Faculty of Natural Sciences and Mathematics, University of Maribor, Koro{\v s}ka  cesta 160, SI-2000 Maribor, Slovenia\\
\inst{3}Faculty of Natural Sciences, Szeged University, H-6720 Szeged, Hungary}


\abstract{Evolution of cooperation in the prisoner's dilemma game is studied where initially all players are linked via a regular graph, having four neighbors each. Simultaneously with the strategy evolution, players are allowed to make new connections and thus permanently extend their neighborhoods, provided they have been successful in passing their strategy to the opponents. We show that this simple coevolutionary rule shifts the survival barrier of cooperators towards high temptations to defect and results in highly heterogeneous interaction networks with an exponential fit best characterizing their degree distributions. In particular, there exist an optimal maximal degree for the promotion of cooperation, warranting the best exchange of information between influential players.}

\begin{document}

\maketitle

\section{Introduction}
Complex networks, and scale-free networks in particular, have been identified as potent promoters of cooperation in all major types of social dilemmas \cite{santos-prl}. Especially the prisoner's dilemma, being one of the most widely applicable games \cite{axelrod}, as well as the snowdrift and ultimatum games, have thus far been studied on diluted \cite{nowak-ijbc, vainstein-pre} and hierarchical networks \cite{vukov-pre}, random graphs \cite{duran-pd}, small-world \cite{abramson-pre, wu-cpl, kuperman-epjb} and real empirical networks \cite{holme-prex}, as well as games on graphs in general \cite{szabo-pr}. Arguably the most important feature of complex networks responsible for the promotion of cooperation is the heterogeneous linkage of participating players, which is due to large differences in the degrees of constitutive nodes. Indeed, studies have shown that masking the heterogeneity via the introduction of participation costs or the usage of normalized or effective payoffs \cite{santos-jeb, tomassini-ijmpc, masuda-prsb, szolnoki-pa} eliminates excessive benefits for cooperators and yields results similar to those reported on regular square lattices \cite{nowak-nat, lindgren-pd, szabo-pre, hauert-nat}. Since heterogeneities due to complex interactions networks have proven wildly successful in promoting cooperation, similar characteristics have been introduced also via differences in the influence and teaching activities of players \cite{kim-pre, wu-pre, szolnoki-epl} and social diversity \cite{perc-pre, santos-nat}.

Besides introducing the relevant heterogeneities artificially, recently it has been shown that the latter can emerge also spontaneously as a part of a coevolutionary process accompanying the evolution of strategies. Particularly, in Ref.~\cite{szolnoki-njp} the teaching activity was considered as an evolving property of players, and it has been shown that simple coevolutionary rules may lead to highly heterogeneous distributions of teaching activity from an initially non-preferential setup, which in turn promotes cooperation in social dilemmas such as the prisoner's dilemma or the snowdrift game. Moreover, similar results were obtained by Poncela \textit{et al.} \cite{poncela-plos}, who considered a coevolutionary preferential attachment and growth scheme to generate complex networks on which cooperation can thrive. An important precursor to these studies were works considering the coevolution of strategy and structure \cite{pacheco-prl} as well as random or intentional rewiring procedures \cite{ebel-pre, zimmermann-pre, eguiluz-ajs, perc-njp}, showing that they as well may help to maintain cooperative behavior. Interestingly, similar effects can also be observed if the players are allowed to move on the lattice during the strategy evolution \cite{vainstein-jtb}.

In this letter, we propose a new model based on a simple coevolutionary rule that, alongside the evolution of the cooperative and the defective strategy within the prisoner's dilemma game, entails increasing the neighborhood of players by allowing them to make new permanent connections with the not yet linked neighbors. The only condition necessary to exercise this is a successful pass of the player's strategy to one of its current opponents. Since each reproduction is considered as a statement of success of the donor player at that time, the latter is rewarded by the expansion of its neighborhood. Thus, the basic premise of the proposed coevolutionary rule is that in real social systems the more successful individuals will typically have more associates than the less successful ones. Related to this, we study the impact of different limits with respect to the maximal degree an individual is allowed to obtain during the coevolutionary process, as well as the degree distribution of thereby resulting networks. Starting from a fully homogeneous and non-preferential setup where each player is linked only to its four nearest neighbors on a square lattice, we show that the model can sustain cooperation even at high temptations to defect, and moreover, that the resulting networks are highly heterogeneous with roughly an exponential degree distribution. In addition, we shed light on the observed phenomena by studying the interconnectedness of influential players and the information exchange between them. Our results suggest that `making new friends' is an essential part of the evolutionary process, playing a crucial role by the sustainability of cooperation in environments prone to defection.

The remainder of this letter is organized as follows. First, we describe the prisoner's dilemma game and the protocol for the coevolution of neighborhoods. Next we present the results, whereas lastly we summarize and discuss their implications.

\section{Game definitions and setup}
We consider an evolutionary prisoner's dilemma game with cooperation and defection as the two competing strategies. The game is characterized by the temptation to defect $T = b$, reward for mutual cooperation $R = 1$, and both the punishment for mutual defection $P$ as well as the suckers payoff $S$ equaling $0$, whereby $1 < b \leq 2$ ensures a proper payoff ranking. Initially, each player $x$ is designated either as a cooperator $(C)$ or defector $(D)$ with equal probability and linked to its four nearest neighbors on a regular $L \times L$ square lattice with periodic boundary conditions, thus having degree $k=4$. This setup warrants that initially all players have equal chances of success, which is crucial for evaluating the impact of the proposed coevolutionary rule. We note, however, that below results are robust against variations in the initial conditions, as well as variations in the parametrization of the prisoner's dilemma game. The evolution of the two strategies is performed in accordance with the Monte Carlo simulation procedure comprising the following elementary steps. First, a randomly selected player $x$ acquires its payoff $p_x$ by playing the game with all its $k_x$ neighbors. Next, one randomly chosen neighbor of $x$, denoted by $y$, also acquires its payoff $p_y$ by playing the game with all its $k_y$ neighbors. Last, if $p_x > p_y$ player $x$ tries to enforce its strategy $s_x$ on player $y$ in accordance with the probability
\begin{equation}
W(s_x \rightarrow s_y)=(p_x-p_y)/b k_q,
\label{eq1}
\end{equation}
where $k_q$ is the largest of $k_x$ and $k_y$. The introduction of $k_q$ is necessary since the degree $k_x$ is presently subject to evolution as well. In particular, each time player $x$ succeeds in passing its strategy to player $y$ the degree $k_x$ is increased by an integer $\Delta k$ according to $k_x \rightarrow k_x + \Delta k$, whereby for simplicity we here use $\Delta k = 1$. Practically, the increase of degree $k_x$ is realized so that player $x$ establishes a permanent new connection with a not yet connected player which is selected randomly amongst the direct neighbors of the current neighborhood of $x$. Thus, successful players are allowed to grow compact large neighborhoods that are centered around their initial four nearest neighbors. Notably, similar results as will be reported below can be obtained if players extend their neighborhoods via long-range connections, but since we primarily wanted to eschew effects of resulting small-world topologies \cite{abramson-pre} and focus solely on the impact of coevolutionary extending neighborhoods, we here present the results obtained with the former model. The described coevolutionary rule would eventually result in a fully connected graph, which in turn would prevent the survival of cooperation due to the applicability of the well-mixed limit. Accordingly, to abridge the latter effect we introduce $k_{max}$ as the maximal degree a player is allowed to obtain. In fact, the coevolutionary process of making new connections is stopped as soon as the degree $k$ of a single player within the whole population reaches $k_{max}$, whereby this limit prevents the formation of a homogeneous system and will be one of the main parameters to be varied below. Despite being strikingly simple, the proposed protocol for the coevolution of neighborhoods is remarkably robust, delivering conclusive results with respect to the final distribution of $k$ as well as the two competing strategies.

In accordance with the random sequential update, each individual is selected once on average during a full Monte Carlo step (MCS), which consists of repeating the above elementary steps $L^2$ times corresponding to all participating players. Monte Carlo results were obtained on populations comprising $100 \times 100$ to $400 \times 400$ individuals, whereby the stationary fraction of cooperators $\rho_C$ was determined within $10^5$ to $10^6$ MCS after sufficiently long transients were discarded. Moreover, as the coevolutionary process yields highly heterogeneous interaction networks, thus yielding heavily fluctuating outputs, final results were typically averaged over $200$ independent runs for each set of parameter values in order to take into account the stochastic feature of host graph topology resulting from the coevolutionary process.

\section{Results}

\begin{figure}
\scalebox{0.45}[0.45]{\includegraphics{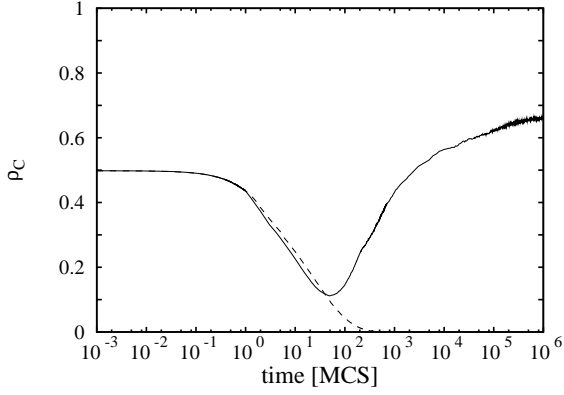}}
\caption{Time evolutions of $\rho_C$ obtained for $b = 1.24$. Dashed line shows results obtained in the absence of the coevolutionary neighborhood growth, while the solid line depicts the outcome of the prisoner's dilemma game if $k_{max} = 50$. Note that the $x$-axis has a logarithmic scale on which fractions of the first full MCS are shown as well for clarity.}
\label{fig1}
\end{figure}

The remarkable impact of the above defined coevolutionary process is demonstrated in Fig.~\ref{fig1}, where the time evolution of $\rho_C$ obtained with and without the inclusion of coevolution is presented. The difference between the two outcomes is evident as the basic version of the game fails to sustain cooperative behavior (dashed line) while the inclusion of the coevolutionary process is able to recover it and maintain respectable $\rho_C = 0.66$ (solid line). Clearly thus, the newly proposed model is able to sustain cooperation in regions of $b$ where the regular square lattice interaction topology fails. However, it is interesting to note that this is true for the final outcome of the game, whereas during the first $100$ MCS it seems that the cooperative behavior will actually fare better without the inclusion of the coevolutionary process. Note that the solid line drops slightly faster than the dashed line during the initial phase of the game. Yet rather surprisingly, the tide then shifts in favor of the cooperative strategy as depicted by the solid line in Fig.~\ref{fig1}; a feature that cannot be observed in case the coevolution of neighborhoods is absent. Figure~\ref{fig1} also suggests that, while the results are essentially robust against variations of initial conditions, the initial density of cooperators shouldn't be too low, as otherwise the promotive impact of coevolution could be missed.

\begin{figure}
\scalebox{0.457}[0.457]{\includegraphics{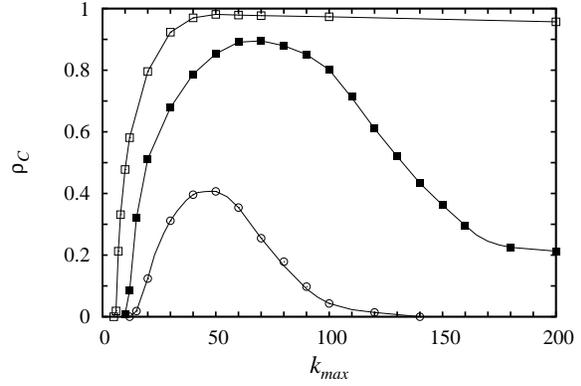}}
\caption{Promotion of cooperation in dependence on $k_{max}$ for $b=1.15$ (open squares), $b=1.2$ (closed squares), and $b=1.28$ (open circles). There exists an optimal value of $k_{max}$ at which $\rho_C$ is maximal. Lines are solely guides to the eye.}
\label{fig2}
\end{figure}

To sharpen the facilitative effect of the coevolutionary rule on the cooperative behavior, we present stationary values of $\rho_C$ in dependence on $k_{max}$ for three different values of $b$ in Fig.~\ref{fig2}. It can be inferred that there exist an optimal maximal degree $k_{max}$ a player is allowed to obtain during the coevolutionary process at which cooperation thrives best. This holds irrespective of $b$, although the optimal values of $k_{max}$ fluctuate between $50$ and $70$. The non-monotonous dependence on $k_{max}$, illustrated in Fig.~\ref{fig2}, is a consequence of the limited support for cooperation offered by the square lattice (results obtained at values of $k_{max}$ equal or close to $4$), and the well-mixed limit that is reached for high $k_{max}$. Note that $k_{max}$ comparable with the system size inevitably lead to a high degree of interconnectedness amongst the players, which is characteristic for a well-mixed population. In between the two extremes, the coevolutionary rule obviously yields a favorable host graph topology for the evolution of cooperation.

\begin{figure}
\scalebox{0.45}[0.45]{\includegraphics{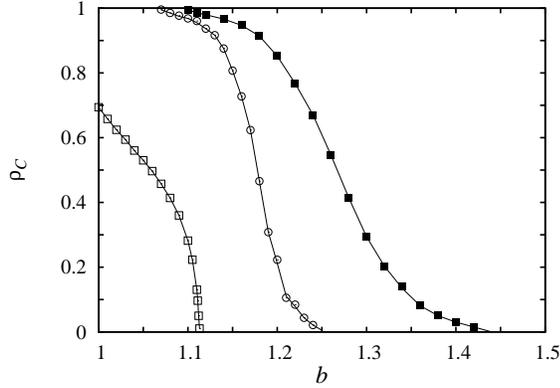}}
\caption{Promotion of cooperation in dependence on $b$ for $k_{max} = 4$ (open squares), $k_{max} = 50$ (closed squares), and $k_{max} = 200$ (open circles). Cooperators are most successful if $k_{max} = 50$, which roughly corresponds to the peak values of $\rho_C$ depicted in Fig.~\ref{fig2}. Lines are solely guides to the eye.}
\label{fig3}
\end{figure}

Figure~\ref{fig3} shows $\rho_C$ in dependence on $b$ for three different values of $k_{max}$, whereby it can be observed that the optimal value of $k_{max} = 50$ is able to sustain some fraction of cooperators almost halfway through the whole span of $b$. By comparison, in the absence of the coevolutionary process (note that $k_{max} = 4$ leaves the initial topology unaltered) the cooperative trait goes extinct at $b = 1.115$. Moreover, large values of $k_{max}$ still yield some advantages for the cooperators, as can be inferred from the $k_{max} = 200$ curve depicted in Fig.~\ref{fig3}, yet increasing $k_{max}$ even further introduces well-mixed-like conditions where the sustainability of cooperation is practically absent even for low values of $b$.

\begin{figure}
\scalebox{0.444}[0.444]{\includegraphics{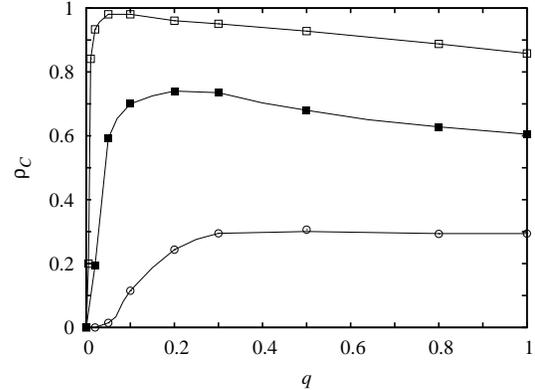}}
\caption{Cooperation level $\rho_C$ in dependence on the time separation between strategy and structure updating $q$ for $b = 1.2$ (open squares), $b = 1.25$ (closed squares), and $b = 1.3$ (open circles). The maximal degree was limited to $k_{max} = 50$ for all three values of $b$. Lines are solely guides to the eye.}
\label{figad}
\end{figure}

Before turning our attention to networks emerging due to the proposed coevolutionary rule, we test above results against the separation of time scales \cite{sanchez-prl}, presently characterizing the evolution of strategies and structure. Thus far, the two time scales were treated as identical since every successful reproduction was followed by an increase in the player's degree. The model can be generalized via a parameter $q$ that determines the probability of degree extension after a successful strategy pass. Evidently, $q=1$ recovers the originally proposed model while decreasing $q$ result in increasingly separated time scales. At $q=0$ the model becomes equivalent to the spatial model without coevolution, hence yielding $\rho_C = 0$ by high $b$, as demonstrated in Fig.~\ref{figad}. An increase in $q$, resulting in a moderately fast yet effective coevolution, is beneficial for cooperation since influential cooperators can then extend their neighborhoods and thus become stronger by collecting higher payoffs already during the coevolutionary process. Oppositely, influential defectors become weaker as their defecting neighborhoods grow, which ultimately results in the highest cooperation levels at intermediate $q$. However, further increasing $q$ can generate a slight downward trend of $\rho_C$ because the influential cooperators cannot take full advantage of their newly acquired neighbors within the short time between consecutive building steps, and thus defectors can gain a slight yet permanent advantage. The moderate decrease in $\rho_C$ due to the too fast network evolution if compared to the strategy evolution is, however, virtually absent by very high $b$, since then the dominating feature is the final heterogeneous network topology rather than initial fights of dominance. For simplicity, and to preserve comparability with the results in the first three figures, we will continue to use $q=1$ in what follows.

\begin{figure}
\scalebox{0.45}[0.45]{\includegraphics{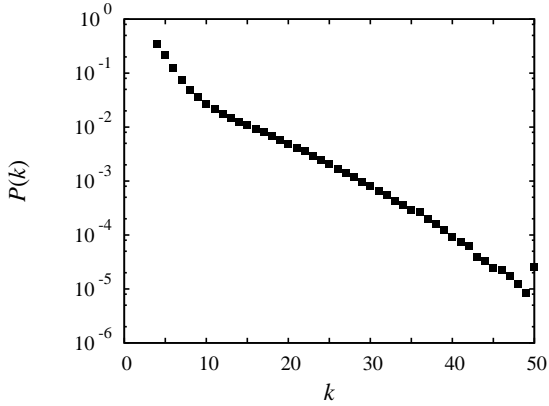}}
\caption{Final distribution of degree $P(k)$ in the studied prisoner's dilemma game obtained for $b=1.26$ via $k_{max} = 50$. Note that the $y$ axis has a logarithmic scale to clearly reveal the heterogeneous outlay of $P(k)$.}
\label{fig4}
\end{figure}

Next, we examine properties of networks resulting from the coevolutionary process. As heterogeneity is the most important property favoring cooperation, we first focus on the degree distribution $P(k)$. Given the fact that substantial promotion of cooperation was in the past often associated with strongly heterogeneous states, either in form of the host network \cite{santos-prl} or social diversity \cite{perc-pre, santos-nat}, it is reasonable to expect that $P(k)$ will exhibit similar features as well. Results presented in Fig.~\ref{fig4} clearly attest to this expectation as the semi-log plot of the distribution reveals a highly heterogeneous outlay of $P(k)$ that can be most accurately described by an exponential fit. The latter feature is crucial for the fortified facilitative effect on cooperation outlined in Figs.~\ref{fig2} and \ref{fig3}, in particular since it incubates cooperative clusters around individuals with high $k$, as described previously in \cite{santos-prl} and reviewed in \cite{szabo-pr}. Contrary, since the positive feedback of the imitating environment is not associated with influential defectors they therefore fail to survive even if temptations to defect are large. As already noted, a similar behavior underlies the cooperation-facilitating mechanism reported for the scale-free network where players with the largest connectivity (presently equivalent to those having $k$ close to $k_{max}$) also act as robust sources of cooperation in the prisoner's dilemma game. Noteworthy since it is related to time courses presented in Fig.~\ref{fig1}, before the heterogeneous network topology fully evolves, defectors temporarily thrive since they gain a larger base of neighbors to exploit. Once, however, the prime spots of the evolved network are overtaken by cooperators the defectors start loosing ground fast, which explains the initial drop and the subsequent recovery of cooperative behavior depicted by the solid line in Fig.~\ref{fig1}. The presently reported spontaneous emergence of the heterogeneous distribution of degree from an initially non-preferential state within the framework of evolutionary game theory suggests that even very simple coevolutionary rules might lead to a strong segregation amongst participating players, which is arguably advantageous for flourishing cooperative states. We argue that the core mechanism responsible for the emergence of heterogeneity in the degree distribution presented in Fig.~\ref{fig4} can be related to the growth and preferential attachment mechanism proposed by Barab\'{a}si and Albert \cite{barabasi-sci}. In particular, our model incorporates preferential attachment in that the probability of increasing the degree is larger by players that have had successful reproductions in the past since they are more likely to reproduce in the future. Obviously, however, our model does not incorporate growth since players are not added in time. Nevertheless, since the evolution is halted by a given $k_{max}$, preferential attachment alone can still lead to highly heterogeneous but not scale-free distributions \cite{barabasi-sci}.

\begin{figure}
\scalebox{0.395}[0.395]{\includegraphics{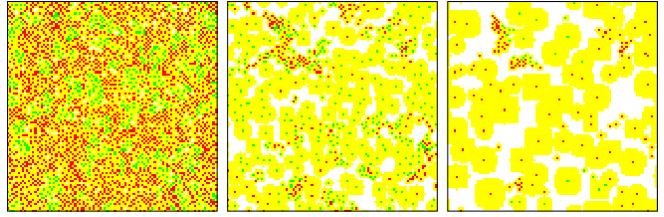}}
\caption{Snapshots of typical distributions of players on a $100 \times 100$  grid, obtained for $k_{max} = 14$ (left panel), $k_{max} = 50$ (middle panel) and $k_{max} = 200$ (right panel) at $b=1.2$. Red and green are influential players (see text for details) in defector and cooperator states, respectively, while yellow are all the direct neighbors of the depicted influential players. If a player is neither influential nor belonging to a neighborhood of an influential player it is marked white.}
\label{fig5}
\end{figure}

One may argue, however, that similar highly heterogeneous degree distributions can be obtained at higher $k_{max}$ as well, yet the promotion of cooperation is then still moderate, as demonstrated in Fig.~\ref{fig2}. This observation highlights that the heterogeneous distribution itself is not a sufficient condition for ample levels of cooperation at high temptations to defect. To uncover the additional decisive feature of resulting networks by different $k_{max}$, we study the overlap of neighborhoods of the so-called influential players, whereby a player is designated as influential if it has the highest degree among any other players that can adopt the strategy from the influential player via an elementary process. Figure~\ref{fig5} shows typical distributions of influential players, which are denoted either green (cooperators) or red (defectors) depending on their strategy. In addition, their neighborhoods, formed by those directly linked to the influential players, are depicted yellow. The distributions are plotted for different values of $k_{max}$ but for an identical temptation to defect equalling $b=1.2$. At small $k_{max}$ there exist many influential players with small neighborhoods surrounding them, yet the outlay is virtually homogenous, and thus the promotion of cooperation is not notably enhanced if compared to the square lattice alone. Around the optimal $k_{max}$, however, influential players become fewer, and their neighborhoods larger. Importantly though, the overlap between their neighborhoods is still remarkable, which is crucial as it enables influential cooperators to overtake influential defectors as soon as the latter weaken their neighborhoods. As we will show next, this effective information transfer between the influential players is crucial for the feedback mechanism to work. At higher $k_{max}$ influential players become rarer still, and their neighborhoods grow further, yet crucially, the overlap between them vanishes, thus hindering influential cooperators to overtake influential defectors. In sum, defectors are virtually undisturbed in exploiting their large neighborhoods, ergo leading to a population in which defection is widespread.

\begin{figure}
\scalebox{0.45}[0.45]{\includegraphics{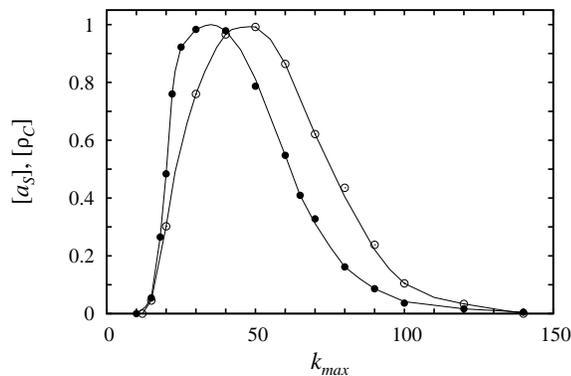}}
\caption{Frequency of strategy adoptions between influential players $a_S$ (closed circles) and the cooperation level $\rho_C$ (open circles) in dependence on $k_{max}$ for $b=1.28$. Both quantities are normalized by their maximal values for better comparisons, and are therefore decorated by square brackets on the corresponding axis label. Lines are solely guides to the eye.}
\label{fig6}
\end{figure}

The impact of the above discussed topological feature can be studied directly by measuring the information transfer between influential players, which we realize via $a_S$ quantifying the frequency of strategy adoptions between influential players in the stationary state of the prisoner's dilemma game for each value of $k_{max}$. Figure~\ref{fig6} features the results, and in addition, shows the cooperation level for the sake of comparison. Note that both $a_S$ and $\rho_C$ are depicted normalized with their maximal values (denoted as $[a_S]$ and $[\rho_C]$ on the vertical axis), yet the outlay of the curves thereby remains unaltered. At values of $k_{max}$ that are comparable to the initial degree of all participating players the neighborhood size of influential individuals is small. Thus, they cannot communicate efficiently with one another, which ultimately results in low values of $a_S$. As the maximally attainable degree limit increases the average neighborhood size of influential players grows as well. Consequently, direct strategy adoptions between them become more frequent, and most crucially, strong influential cooperative players can overtake weakened influential defectors, in turn allowing the feedback mechanism to blossom in the intermediate region of $k_{max}$. Contrary, when $k_{max}$ exceeds the optimal value some players become too influential and grow well separated neighborhoods (see the right panel of Fig.~\ref{fig5}), which hampers the information exchange between them so that influential defectors can prevail for prolonged periods of time despite of their weak posture inflicted by the defecting neighborhoods. Indeed, results presented in Fig.~\ref{fig6} show that the level of information exchange, quantified via $a_S$, and the global cooperation level in a heterogeneous environment are strongly bound to one another, following very similar patterns depending on $k_{max}$, thus validating our reasoning. Moreover, it is worth noting that our explanation is in agreement with a previous observation of Rong \textit{et al.} \cite{rong-pre07}, who detected a fall of overall cooperation levels in the prisoner's dilemma game on scale-free networks following disassortative mixing, which also results in an enhanced isolation of hubs.

\section{Summary}
We have demonstrated that the introduction of a very simple coevolutionary process to the spatial prisoner's dilemma game markedly improves survival chances of cooperators in highly defection-prone environments, and moreover, enhances their overall dominance at moderate temptations to defect. Most notably, the underlying mechanism behind the reported promotion of cooperation is routed in the resulting highly heterogeneous network structure which emerges spontaneously from a non-preferential setup following a simple coevolutionary rule that indirectly promotes players that are able to pass their strategy by allowing them to extend their neighborhoods via new connections to not yet linked players. Moreover, we have shown that the newly introduced coevolutionary rule yields optimal results if limited by a maximal degree the most influential player is allowed to obtain. If this limit is surpassed, the detrimental impact on the evolution of cooperation sets in due to the decreasing overlap of neighborhoods of the influential players, which enables defectors to reign in isolation from potentially stronger influential cooperators. The success driven increase of degree indirectly introduces a preferential attachment mechanism into the model, which combined with the limiting $k_{max}$, results in heterogeneous degree distributions.

In sum, presented results confirm that the presence of influential leaders is advantageous for cooperation, and more importantly, that a simple `making new friends' coevolutionary rule may bring about just the appropriate diversity between participating players if appropriately timed. The coevolutionary model demonstrates how influential leaders can evolve from an initially non-preferential state, and that it is optimal for cooperation if their overall density remains bounded to intermediate levels.

\acknowledgments
Discussions with Gy{\"o}rgy Szab{\'o} are gratefully acknowledged.

\end{document}